%% file: main.tex
\documentclass[conference,compsoc]{IEEEtran}

\ifCLASSOPTIONcompsoc
  \usepackage[nocompress]{cite}
\else
  \usepackage{cite}
\fi

\input{macros}
\begin{document}

\title{\paperTitle}

\input{authors}


\maketitle
\pagestyle{plain}
\IEEEpeerreviewmaketitle

\input{content}

\appendices

\bibliographystyle{IEEEtran}
\bibliography{ref}

\appendices
\input{sections/appendix}

\end{document}

%% file: macros.tex
\usepackage{booktabs}    
\usepackage{array}       
\usepackage{tabularx}
\usepackage{enumitem}
\usepackage{longtable}
\usepackage{tcolorbox}
\usepackage[table]{xcolor}
\usepackage{tikz}
\usepackage[utf8]{inputenc}
\usepackage{textgreek}
\usepackage{algorithm}
\usepackage{algpseudocode}
\usepackage{textcomp}
\usepackage{caption}
\usepackage{pifont}
\usepackage{xcolor}
\usepackage{adjustbox}
\usepackage{xspace}
\usepackage{multirow}
\usepackage{mdframed}

\usepackage{minted}  
\usepackage{tcolorbox} 
\tcbuselibrary{skins,breakable,minted}
\usepackage{url}
\usetikzlibrary{shapes.geometric, arrows, positioning, calc, arrows.meta}

\usepackage{amsmath}
\usepackage{amssymb}

\newcommand{\toolname}{\textsc{VeriPort}\xspace}
\newcommand{\autoresearch}{\textsc{AutoResearch}\xspace}
\newcommand{\autopoc}{\textsc{AutoPoc}\xspace}
\newcommand{\autotester}{\textsc{AutoTester}\xspace}
\newcommand{\autopatch}{\textsc{AutoPatch}\xspace}

\newcommand{\cve}{CVE\xspace}
\newcommand{\poc}{PoC\xspace}
\newcommand{\benchmarkAmount}{393\xspace}
\newcommand{\backportBenchmarkTotalAmount}{128\xspace}
\newcommand{\pocs}{PoCs\xspace}
\newcommand{\benchname}{CVEPatchBench\xspace}

\newcommand{\foundghsaFP}{2{,}100\xspace}
\newcommand{\foundghsaFN}{127\xspace}
\newcommand{\foundghsaAdvisories}{92\xspace}
\newcommand{\mergedghsaFP}{387\xspace}
\newcommand{\mergedghsaFN}{81\xspace}
\newcommand{\mergedghsaPRs}{23\xspace}

\newif\ifsubmission
\submissionfalse
\newcommand{\companydesc}{software supply chain security company\xspace}
\ifsubmission
  \newcommand{\company}{a \companydesc}   
  \newcommand{\companyappos}{}            
\else
  \newcommand{\company}{Socket\xspace~\cite{SOCKET}}
  \newcommand{\companyappos}{, a \companydesc}
\fi

\definecolor{VerifyLightRed}{RGB}{255, 104, 103}

\newcommand{\backportReliabilityTool}{95.3\%\xspace}
\newcommand{\backportReliabilityBaseline}{72.6\%\xspace}

\newcommand{\passhat}[1]{\ensuremath{\mathrm{pass}^{#1}}}
\newcommand{\passat}[1]{\ensuremath{\mathrm{pass}@#1}}

\usepackage{listings}

\newcommand{\circletext}[2]{%
  \tikz[baseline=(char.base)]{
    \node[shape=circle, draw=#1, fill=#1, text=white, inner sep=0.5pt] (char) {\footnotesize\bfseries #2};
  }%
}

\newcommand{\squaretext}[2]{%
  \tikz[baseline=(char.base)]{
    \node[shape=rectangle, rounded corners=1.2pt, draw=#1, fill=#1, text=white, inner sep=2pt] (char) {\footnotesize\bfseries #2};
  }%
}

\newcommand{\myparagraph}[1]{\vspace{0.05em}\noindent\textbf{#1}.}

\newcommand{\myparagraphn}[1]{\vspace{0.1em}\noindent\textbf{#1}} 

\newtcblisting{codeblock}[3][]{%
  listing engine=minted,
  listing only,              
  minted language=#2,
  minted options = {
    breaklines,
    autogobble,
    linenos,
    numbersep=8pt,
    fontsize=\small,
    tabsize=2,
    #1
  },
  colback=gray!2,
  colframe=gray!60,
  coltitle=black,
  title={#3},
  left=8pt, right=8pt, top=6pt, bottom=6pt,
  boxrule=0.6pt, arc=2mm,
  enhanced, breakable
}

\newcommand{\yesIconColor}{green!50!black}
\newcommand{\noIconColor}{red!70!black}
\newcommand{\yesIconSize}{0.8ex}
\DeclareRobustCommand{\no}{\textcolor{\noIconColor}{\ding{55}}}
\DeclareRobustCommand{\yes}{\textcolor{\yesIconColor}{\ding{51}}}
\DeclareRobustCommand{\partialyes}{%
  \tikz[baseline=-0.55ex]{
    \draw[line width=0.12ex] (0,0) circle (\yesIconSize);
    \fill[\yesIconColor] (0,0) -- (0,\yesIconSize) arc (90:270:\yesIconSize) -- cycle;
  }%
}

\definecolor{custompurple}{HTML}{7F00FF}
\definecolor{customred}{HTML}{CC0001}
\definecolor{customgreen}{HTML}{038A00}
\definecolor{customorange}{HTML}{FE8701}
\definecolor{custombluedark}{HTML}{014A94}
\definecolor{custombluelight}{HTML}{007FFF}

\newcommand{\paperTitle}{\toolname: Automated and Verified Patch Backporting at Scale}

\newcounter{finding}
\newcommand{\findingname}{Finding}

\mdfdefinestyle{findingstyle}{%
  linewidth=0.8pt,
  linecolor=black!40,
  backgroundcolor=black!5,
  roundcorner=6pt,
  shadow=true,
  shadowsize=3pt,
  shadowcolor=black!20,
  innertopmargin=7pt,
  innerbottommargin=7pt,
  innerleftmargin=9pt,
  innerrightmargin=9pt,
  skipabove=7pt,
  skipbelow=7pt,
}

\NewDocumentCommand{\finding}{o m}{%
  \refstepcounter{finding}%
  \IfValueT{#1}{\label{#1}}%
  \begin{mdframed}[style=findingstyle]%
    \textbf{\findingname~\thefinding.}\quad\textit{#2}%
  \end{mdframed}%
}

\newcommand{\takeawayname}{Takeaway}

\NewDocumentCommand{\takeaway}{m}{%
  \begin{mdframed}[style=findingstyle]%
    \textbf{\takeawayname.}\quad\textit{#1}%
  \end{mdframed}%
}

\newcommand{\defaultBackportbench}{\emph{Patch Adaptation}\xspace}

\newcommand{\repoBackportbench}{\emph{Patch Isolation and Adaptation}\xspace}

\usepackage{amsmath}
\usepackage{amsthm}

\theoremstyle{definition}
\newtheorem{defn}{Definition}

\let\oldproofname=\proofname
\renewcommand{\proofname}{\rm\bf{\oldproofname}}

\usepackage{mathtools}

\usepackage[hidelinks]{hyperref}
\usepackage{cleveref}
\crefname{section}{Section}{Sections}
\Crefname{section}{Section}{Sections}
\crefname{subsection}{Section}{Sections}
\Crefname{subsection}{Section}{Sections}
\crefname{subsubsection}{Section}{Sections}
\Crefname{subsubsection}{Section}{Sections}

\newcommand{\msp}{\ensuremath{\mathcal{M}}\xspace} 

%% file: authors.tex
\makeatletter
\newcommand{\linebreakand}{%
  \end{@IEEEauthorhalign}
  \hfill\mbox{}\par
  \mbox{}\hfill\begin{@IEEEauthorhalign}
}
\makeatother

\IEEEoverridecommandlockouts


\author{
\IEEEauthorblockN{
\href{https://orcid.org/0009-0006-8915-5923}{Jonah Ghebremichael}\IEEEauthorrefmark{2}\textsuperscript{*},
\href{https://orcid.org/0000-0003-2608-8576}{Wenxin Jiang}\IEEEauthorrefmark{3}\textsuperscript{*},
\href{https://orcid.org/0009-0006-8771-2503}{Mikola Lysenko}\IEEEauthorrefmark{3},
\href{https://orcid.org/0009-0006-6470-0884}{Benjamin Barslev Nielsen}\IEEEauthorrefmark{3}, \\
\href{https://orcid.org/0000-0002-3043-8092}{William Enck}\IEEEauthorrefmark{2},
\href{https://orcid.org/0000-0002-8839-8521}{Alexandros Kapravelos}\IEEEauthorrefmark{2}\IEEEauthorrefmark{3}
}
\IEEEauthorblockA{\IEEEauthorrefmark{2}\textit{North Carolina State University} --- \{jghebre, whenck, akaprav\}@ncsu.edu}
\IEEEauthorblockA{\IEEEauthorrefmark{3}\textit{Socket Inc.} --- \{wenxin, mik, barslev, alexandros\}@socket.dev}
\thanks{\textsuperscript{*}These authors contributed equally to this work.}
}

%% file: content.tex
\input{sections/abstract}

\section{Introduction}
\label{sec:Intro}

\input{sections/intro}

\section{Problem and Related Work}
\label{sec:problem}
\input{sections/motivation}

\section{Overview}
\label{sec:OurApproach}
\input{sections/approach}

\section{\toolname}
\label{sec:methodology}
\input{sections/methodology}

\section{Evaluation}
\label{sec:Eval}
\input{sections/eval}

\section{Production Deployment}
\label{sec:Production}

\input{sections/deployment}

\section{Conclusion}
\label{sec:Conclusion}
\input{sections/conclusion}

\section*{Acknowledgements}
\label{sec:acknowledge}
\input{sections/acknowledgements}



%% file: sections/abstract.tex
\begin{abstract}

One of the key challenges for securing the software supply chain is addressing known vulnerabilities in third-party open-source dependencies.
Security patches are frequently only available for the latest version of a dependency, leaving developers with the choice of either upgrading to the latest version (risking breaking changes) or manually backporting the security fix.
Prior work backports to a single version that must be specified in advance
and does not produce sufficient evidence to demonstrate that their patches block exploitation and preserve functionality.
%
In this paper, we present \toolname, an end-to-end agentic system that scalably backports a patch for a given vulnerability advisory to every affected version of the package.
For each backport, \toolname builds a chain of evidence 
to confirm that the patch blocks exploitation and preserves intended behavior.
\toolname reliably resolves 95.3\% of  \backportBenchmarkTotalAmount\ backporting tasks in BackportBench, outperforming the best existing solution (Claude Code) by 22.7 percentage points.
We further deployed \toolname on 169 high- and critical-severity CVEs and have generated over 5,000 verified backported patches.
Moreover, \toolname's value extends beyond simply backporting patches.
It uncovered \foundghsaFP versions incorrectly reported as affected and \foundghsaFN previously unidentified vulnerable versions across \foundghsaAdvisories advisories, and \mergedghsaPRs advisories have since been corrected upstream by removing \mergedghsaFP versions and adding \mergedghsaFN.
\end{abstract}

%% file: sections/intro.tex
Software vulnerabilities are being disclosed at an unprecedented rate~\cite{gamblin2025cve, gamblin2024cve, nvd}.
This pace is accelerating  as frontier Large Language Models (LLMs) become capable of autonomously discovering zero-day vulnerabilities in widely deployed software~\cite{MYTHOS}.
Most vulnerabilities affect more than one major version of a package, but a fix is usually only published for the latest release~\cite{Decan2022Back}.
Developers using an older version cannot always adopt that fix by upgrading, since a newer release often introduces breaking changes~\cite{bogart2021breaking, UpgradeMaven}.
The alternative is \textit{backporting}, which adapts the upstream fix to apply to an older version.

Backporting remains a primarily human-driven task that requires an engineer to understand the vulnerability, reconstruct the upstream fix against code that has diverged from the patched release, and confirm that the result still blocks the exploit without breaking existing behavior.
The effort is significant enough that projects like the Linux kernel maintain dedicated teams to backport fixes to specific older releases~\cite{LINUXSTABLETEAMS}.
Most projects cannot sustain this investment, making existing backports scarce.
To close this gap, a line of research has emerged to automate the backporting process.

Early automated backporting tools adapt a fix by matching it against a predefined model, whether as syntactic patterns or as program analysis that locates the corresponding code in the target~\cite{FixMorph,Coccinelle,TSBPORT,SKYPORT}.
These tools work well when the divergence between versions follows well-defined patterns, but they cannot generalize to structural changes that fall outside their predefined rules.
Each of these tools requires as input a \emph{minimal security patch} (MSP): a patch containing only the changes that fix the vulnerability.
An MSP rarely exists in published advisories, and even when a vulnerability-fixing commit (VFC) is linked, it can tangle the fix with unrelated changes; therefore extracting one is a hard problem in its own right~\cite{VFCFINDER,FIXSEEKER}.
Moreover, these automated backporting tools judge a backport by its resemblance to a reference patch or by static checks, rather than confirming it blocks Proof-of-Concept (PoC) exploits and preserves existing functionality~\cite{BackPortBench}.
Together these limitations confine existing tools to a narrow slice of advisories, and require significant human effort for verification.

\myparagraph{Insights}
Through our experience backporting packages, we observe that an automated backporting tool must perform four core steps.
It must  \circletext{gray}{1}~gather rich context about the affected package, the vulnerability, and the upstream fix;  \circletext{gray}{2}~generate PoCs;   \circletext{gray}{3}~generate regression suites; and  \circletext{gray}{4}~isolate the MSP from non-security changes, adapt it to affected versions, and validate the proposed patch via generated tests.
Each of these steps is hard because it requires reasoning about code semantics rather than matching against fixed patterns.
A system must separate the security fix from the unrelated changes around it, follow that fix through versions where code has been restructured, and craft inputs that genuinely trigger the vulnerability.
Modern LLMs have grown capable of exactly this kind of code reasoning, which makes them a natural foundation for this workflow~\cite{SWEAGENT}.

Recent systems have begun to apply LLMs to this task, adapting the fix directly and reducing much of the rigidity of earlier rule-based methods~\cite{Mystique,PORTGPT}.
They generalize further than prior approaches but still require an MSP and leave the same verification gaps.
Using an LLM also introduces failure modes of its own.
A model can hallucinate or fabricate information, producing a convincing but wrong result~\cite{LLMHallucinations}.
A model also has a fixed context limit, and approaching that limit degrades performance~\cite{CONTEXT-LIMITS}.
A single advisory spans dozens of affected versions, so regenerating every artifact per version multiplies cost and gives each failure a fresh chance to recur.

\myparagraph{Our Approach}
We present \toolname, an end-to-end agentic system that takes a vulnerability advisory as input and backports the corresponding fix into a minimal, verified patch for each affected version.
Our key insight is that each of the three LLM limitations above yields a corresponding design principle.
First, \toolname never accepts a claim on a single model's output alone, pairing fresh-context critic agents with deterministic checks.
Second, \toolname decomposes backporting into stages, and each stage into narrowly scoped agents, so that every task fits into a bounded context.
Third, \toolname generates and verifies each artifact once and then adapts it to lower affected versions, regenerating from scratch only when the code drift between versions is too large to adapt across.

\toolname realizes these principles in four stages.
\circletext{gray}{1}~\autoresearch reads the advisory and produces a grounded analysis of the package, the vulnerability, the upstream fix, and how the code varies across affected versions.
\circletext{gray}{2}~\autopoc builds a \textit{vulnerability oracle} for each affected version: an exploit that succeeds on a vulnerable version and is blocked once the patch is applied.
\circletext{gray}{3}~\autotester builds a \textit{functionality oracle} for each affected version: a regression suite that exercises intended functionality on both the vulnerable and patched versions.
\circletext{gray}{4}~\autopatch isolates the MSP, adapts it for downstream versions, and accepts a backport only when its vulnerability and functionality oracles pass.
Together these stages form a chain of evidence, each piece built independently of the patch.
A defective backport is accepted only by evading every check at once.

\myparagraph{Results}
We compare \toolname against four baselines on the npm and PyPI tasks of BackportBench~\cite{BackPortBench}, a multilingual benchmark of backporting tasks.
The baselines pair two generic agents, Claude Code~\cite{CLAUDECODE} and mini-swe-agent~\cite{SWEAGENT}, with two patching tools, MagentLess~\cite{MAGENTLESS} and the state-of-the-art agentic backporting system PortGPT~\cite{PORTGPT}.
We run each tool ten times per task and evaluate two settings, one that provides the upstream MSP and one that withholds it.
With the MSP provided, the generic agents solve nearly every task on a single run, and \toolname leads on every metric.
Withholding the MSP costs Claude Code 13.2 percentage points of single-run success and mini-swe-agent 17.3, while \toolname falls only 0.8.
Without the MSP, \toolname reliably solves \backportReliabilityTool of tasks on all ten runs, against \backportReliabilityBaseline for Claude Code, the strongest baseline.
\toolname's worst case across the ten runs exceeds every baseline's best case, and it is the only tool in either setting to solve every task within ten runs.

We further compare \toolname to Claude Code on \benchname, a benchmark we introduce to evaluate backporting across the npm ecosystem, consisting of 393 backporting tasks.
Running each tool three times per task, \toolname reliably solves 91.3\% of tasks on all three runs against 71.1\% for Claude Code.
Its worst case exceeds Claude Code's best case, by 10.7 percentage points, and its pass rate varies by only 2.3 points across runs against Claude Code's 9.5.
We perform an ablation study to isolate the contributions of \autopoc and \autotester.
Removing both verification stages lowers \toolname performance by 9.8 points, showing their importance.

We make the following contributions:
\begin{itemize}
    \item \textbf{End-to-end verified backporting.}
    We present \toolname, the first system to backport an advisory's fix across its entire affected range, with each backport verified by a PoC the patch blocks and a regression suite it passes.
    \item \textbf{A new benchmark.}
    We present \benchname~\cite{cvepatchbench-public},
    \benchmarkAmount npm backporting tasks, each with a containerized environment, a working \poc, a regression suite, and a reference backport that blocks the \poc and passes the suite.
    \item \textbf{Real-world impact.}
    We deployed \toolname in production at \company\companyappos, generating over 5,000 verified backported patches~\cite{PUBLISHEDPATCHES} for 169 high- and critical-severity CVEs, each reviewed by a security engineer before release.

\end{itemize}

\noindent Since \toolname runs an exploit against every release in an advisory's affected range, it observes when a listed version is not exploitable.
\toolname also probes outward from the edges of the affected version range to find unlisted versions that are exploitable.
In deployment, \toolname has flagged \foundghsaFP versions incorrectly listed as affected and \foundghsaFN missed vulnerable versions across \foundghsaAdvisories advisories.
We reviewed each flagged error and reported confirmed corrections upstream.
At the time of writing, corrections to \mergedghsaPRs of these advisories have been merged into the GitHub Advisory Database, removing \mergedghsaFP of the incorrectly listed versions and adding \mergedghsaFN of the missing versions.

%% file: sections/motivation.tex
\input{figures/tool-comparison}

By treating backporting primarily as a code transformation problem, prior work overlooks a key property required in practice: \emph{reliability}.
A reliable backport of a patch 
(a)~fixes the vulnerability and 
(b)~preserves existing behavior.
However, in order to test reliability, the backporting tool requires an understanding of the package, vulnerability, and fix.
Reliably backporting patches requires overcoming four challenges presented in the remainder of this section.

\subsection{Challenge 1: Data Quality}
\label{sec:challenge-data-quality}

Security advisories are a practical source for context about the package, vulnerability, and fix.
They often contain version ranges, fix references, \pocs{}, and external write-ups.
However, they are also known to be noisy and therefore require validation before use in backporting.

\myparagraph{Core Difficulty}
Vulnerability-fixing commits (VFCs) are missing from 63\% of vulnerability database reports~\cite{VFCFINDER} and from roughly 93\% of \cve{}s more broadly~\cite{nguyen2025mapping}.
Even when present, 70\% of VFCs span multiple correlated hunks and are often bundled with unrelated code modifications~\cite{FIXSEEKER, TANGLEDCOMMITS}.
Affected-version metadata is similarly unreliable. Only 59.8\% of vulnerability reports match NVD's standardized version entries~\cite{dong2019towards}, and over half of examined \cve{}s contain spurious affected-version information~\cite{bao2022vszz}.
References are not authoritative either; 91.8\% of GitHub Security Advisories pass through the review pipeline without editorial review~\cite{segal2026ghsa}.

\myparagraph{Existing approaches and gaps}
Existing automated backporting tools such as FixMorph~\cite{FixMorph}, TSBPORT~\cite{TSBPORT}, SKYPORT~\cite{SKYPORT}, and PortGPT~\cite{PORTGPT}  assume that a minimal security patch (MSP) or known upstream security fix has already been identified.
Mystique~\cite{Mystique} further requires the vulnerable function as input. 
VFC-linking and silent-fix detection tools~\cite{VFCFINDER,FIXSEEKER,PATCHSEEKER} narrow this gap by ranking candidate fixing commits or classifying commits as vulnerability fixes, but their outputs remain imperfect and commit-level.
Hence they are unable to separate the MSP from the VFC.
More recent work~\cite{DIFFPLOIT} has begun to test affected version ranges directly using adapted exploits.
This information confirms vulnerability presence but not fix correctness, and so is insufficient for backporting on its own.

\subsection{Challenge 2: Exploit Validation}
\label{sec:challenge-exploit-val}

Since vulnerable code can shift between versions, it is insufficient to simply check if a backported patch applies cleanly.
A reliable backporting tool requires a vulnerability oracle to determine if the vulnerability is actually fixed.
PoCs commonly serve as such oracles~\cite{PORTGPT, CVEGENIE}.

\myparagraph{Core Difficulty}
A working \poc{} must identify an attacker-controlled entry point, construct input that reaches the vulnerable behavior, rebuild the environment where the bug manifests, and define a success condition that cannot be satisfied by an unrelated crash or fabricated output.
These requirements vary by vulnerability class and runtime context. Reproducing the bug may require non-default configurations or deployment settings~\cite{Ruan_2024, CVEGENIE}.
Security advisories often omit executable \pocs{} and describe the bug informally~\cite{POCGEN}.

\myparagraph{Existing approaches and gaps}
Prior automated backporting tools~\cite{FixMorph,TSBPORT,SKYPORT,Mystique,PORTGPT} evaluate candidate patches by similarity to developer-created backports, static properties, or supplied tests.
A more robust approach is to generate and execute a \poc for the specific version of the code that the patch is created for.
Recent work has considered LLM-driven \poc{} generation~\cite{POCGEN, CVEGENIE}, but exploit construction remains unreliable in practice.

\subsection{Challenge 3: Regression Validation}
\label{sec:challenge-regression-val}

A backported patch requires a functionality oracle to ensure it does not break intended functionality.
Project regression tests can help create such an oracle; however, they are not ubiquitous. 
Automatically generating regression tests
is particularly subtle, because the tests need to differentiate intended behavior from unintended (vulnerable) behavior.

\myparagraph{Core Difficulty}
More than 90\% of analyzed npm releases ship without test code, and reconstructed suites often provide limited coverage~\cite{sun2021automatically}.
Further, we found naive regression tests can include vulnerable behavior.
Therefore, a backporting tool must identify or generate a relevant set of regression tests for a given vulnerability.
Traditional automated generators~\cite{fraser2011evosuite}
infer assertions from current behavior, which can encode buggy or incomplete semantics.
Modern LLM-based generators~\cite{EmpiricalUnitTests,berndt2026flakiness,alshahwan2024automated} use code and natural-language context to infer intent, but still create invalid tests, hallucinated APIs, and flaky assertions.

\myparagraph{Existing approaches and gaps}
Existing backporting tools~\cite{FixMorph,TSBPORT,SKYPORT,Mystique,PORTGPT} rely on developer-created backports, static checks, or supplied test suites rather than generating regression tests for the target version.
General-purpose test generation can exercise code near the fix, but no existing approach determines which pre-patch behavior should survive and which is related to the vulnerability.

\subsection{Challenge 4: Code Divergence}
\label{sec:challenge-code-divergence}

Across versions, files move, functions are renamed or split, APIs change, and the same behavior can be implemented through different control flow.
As divergence accumulates, applying the upstream diff directly can fail, patch the wrong location, or miss supporting changes that are required only in the older version.

\myparagraph{Core Difficulty}
The pervasiveness of breaking changes highlights the core difficulty of backporting patches to older versions.
Nearly one-third of Maven releases introduce at least one breaking change~\cite{Raemaekers2017}.
In npm, 12\% of dependent packages and 14\% of their releases are broken by non-major dependency updates, and 44\% of observed breaking changes occur in minor or patch releases~\cite{VENTURI}.
These measurements understate the pervasiveness of breaking changes: semantic changes can also alter behavior but without changing API signatures~\cite{Zhang2022Sembid}.
Therefore, each affected version may require a different adaptation of the same underlying fix.

\myparagraph{Existing approaches and gaps}
Existing backporting tools make important progress, but they each have key limitations.
FixMorph~\cite{FixMorph} synthesizes transformation rules from the upstream patch and target-version alignment.
This approach works when the required change fits the learned syntactic transformation, but it struggles with larger semantic or structural edits.
SKYPORT~\cite{SKYPORT} improves localization with graph-based program analysis to backport injection-related patches.
However, this approach requires predefined sinks, which does not generalize to all vulnerability classes.
TSBPORT~\cite{TSBPORT} uses program-dependence-graph matching and patch-type-specific migration, but it is limited to pre-defined patterns.
Mystique~\cite{Mystique} replaces fixed rules with a fine-tuned LLM, but it operates at function granularity and assumes the corresponding vulnerable function in the target version is available as input.
PortGPT~\cite{PORTGPT} generalizes further with an agentic workflow and repository tools, but it fails to reason across hunks and its effectiveness drops on complex cases that require prerequisite changes, missing functions, or large structural edits.
These tools show that patch adaptation is feasible, but they leave open the harder setting where one advisory must be adapted across many affected versions whose files, APIs, and code paths have drifted in different ways.

%% file: figures/tool-comparison.tex
\begin{table*}[t!]
\centering
\footnotesize
\setlength{\aboverulesep}{0.2ex}
\setlength{\belowrulesep}{0.2ex}
\caption{\toolname compared to prior automated backporting tools.
\yes~denotes full support, \partialyes~denotes partial support, and \no~denotes no support.}
\label{tab:comparison}
\resizebox{\textwidth}{!}{%
\begin{tabular}{l >{\columncolor{gray!10}}c c c c c c}
\toprule
\textbf{Capability}
& \textbf{\shortstack{\toolname\\(This work)}}
& \shortstack{PortGPT~\cite{PORTGPT}\\(S\&P 26)}
& \shortstack{Mystique~\cite{Mystique}\\(FSE 25)}
& \shortstack{TSBPORT~\cite{TSBPORT}\\(CCS 23)}
& \shortstack{SKYPORT~\cite{SKYPORT}\\(USENIX 22)}
& \shortstack{FixMorph~\cite{FixMorph}\\(ISSTA 21)} \\
\midrule
Vulnerability and fix modeling   & \yes & \partialyes & \partialyes & \partialyes & \partialyes & \no         \\
Minimal-Security-Patch isolation           & \yes & \no         & \no         & \no         & \no         & \no         \\
Patch adaptation    & \yes & \yes        & \yes        & \yes        & \partialyes & \yes        \\
\midrule
Exploit oracle                    & \yes & \partialyes & \no         & \no         & \no         & \no         \\
Regression oracle                 & \yes & \partialyes & \no         & \no         & \no         & \partialyes \\
\midrule
Deterministic                     & \no  & \no         & \no         & \yes        & \yes        & \yes        \\
\bottomrule
\end{tabular}%
}
\end{table*}

%% file: sections/approach.tex
\begin{figure}
    \centering
    \includegraphics[width=1\columnwidth]{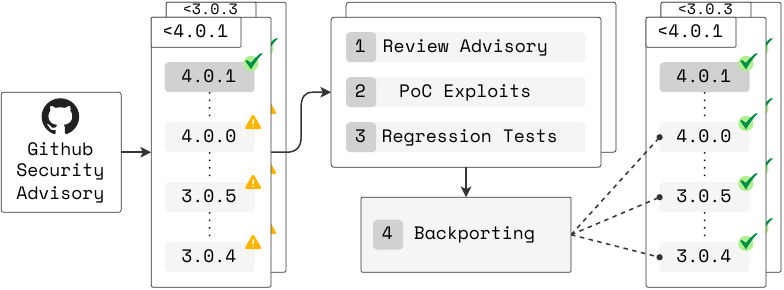}
    \caption{\toolname system overview.}
    \label{fig:system-overview}
\end{figure}

\toolname's goal is to produce a non-breaking security patch for every affected version of a given advisory.
The obstacle is not generating these patches but trusting them.
Modern LLMs can draft a plausible patch for any affected version, but a plausible patch is not a trustworthy one.
A backport that breaks functionality incentivizes users to delay or skip future security updates.
A backport that fails to block the exploit leaves users vulnerable while appearing safe.
We call a patch with either flaw \emph{defective}.
A patch is therefore only useful when paired with evidence that it is not.

\subsection{LLM Limitations}
\label{sec:llm-limitations}
 
Section~\ref{sec:problem} identified four challenges that stand between an advisory and that evidence: data quality, exploit validation, regression validation, and code divergence.
The difficulty in each challenge is largely semantic, not mechanical.
Resolving them requires verifying and enriching advisory data, constructing working exploits, inferring intended behavior, and restructuring a fix across code drift.
LLMs can supply this semantic reasoning, but an LLM's output is a stochastic claim, not trusted evidence.
Three limitations stand between that reasoning and a trustworthy backport.

\myparagraph{Unsupported claims}
An LLM can produce a convincing but wrong result.
It may cite a reference that does not support its claim, mistake an unrelated crash for exploitation, or generate a test that preserves buggy behavior.
Each of these mistakes can lead to a defective patch.

\myparagraph{Bounded context}
A single agent that researches, exploits, tests, and patches accumulates stale assumptions, failed attempts, and irrelevant details in one long context.
The context window imposes a hard limit on how much of this a model can see at once.
Well before that limit, the noise competes with the facts needed for the next decision.

\myparagraph{Compounding at scale}
A single advisory often spans dozens of affected versions, and reviewed npm GitHub advisories average 77 affected versions.
Every artifact costs time and compute to generate.
Generating each artifact independently for every version multiplies that cost by the size of the range and gives the failures above a fresh chance to surface each time.

The rule- and graph-based tools sidestep all three by being deterministic, but suffer in generality (Table~\ref{tab:comparison}).
The LLM-based tools improve in generality, but inherit the same limitations: both accept patches on checks a defective patch can pass, both adapt at a narrow granularity, PortGPT~\cite{PORTGPT} hunk by hunk and Mystique~\cite{Mystique} at the function level, and both repeat the full effort for every affected version.
\toolname mitigates this tradeoff, retaining the generality of stochastic reasoning while recovering trust through evidence.

\subsection{Our Approach}
\toolname turns unreliable semantic reasoning into trustworthy patches through three design principles, one per limitation.

\myparagraph{Evidence before trust}
\toolname never accepts a claim on the model's output alone.
No agent grades its own work, and every claim must pass an independent check before anything downstream depends on it.
Claims that can be executed require execution before they can be trusted.
An exploit must trigger the vulnerability on a vulnerable version and fail on a patched one.
A regression test must pass on both the vulnerable and patched versions.
Claims that cannot be executed are audited by critic agents with fresh context that search for missing evidence, inconsistent reasoning, and invalid assumptions.
When both forms of evidence exist, execution outranks judgment.

\myparagraph{Decompose for independent verification}
\toolname decomposes backporting into stages whose tasks each fit a bounded context.
Each stage hands downstream a structured artifact with a short rationale rather than a conversation transcript.
The same separation is what gives the first principle its force.
The exploit, the tests, and the critics are constructed before the patch exists and independently of it.
A defective patch ships only if it evades all of them at once.

\myparagraph{Solve once, then adapt}
A key insight is that although a security advisory may have many affected versions, these versions share the same vulnerability and differ only in how the surrounding code has drifted.
To take advantage of this sharing, \toolname resolves an advisory's affected version ranges into \emph{backport chains} (Definition~\ref{def:chain}), one per fixed version.
A chain orders the affected versions that one fix must reach, beginning at the \emph{highest affected version}, the affected version closest to that fix.
\toolname generates and verifies every artifact once per chain, at the highest affected version, where divergence from the fix is smallest.
It then adapts the exploit, the tests, and the patch down the chain, re-verifying on every version, so verified work is reused rather than regenerated.
\toolname also maps how the vulnerability-relevant code changes across the versions of each chain, so each adaptation works from the actual differences it must bridge.
Together, these structures reduce each additional version to a small, verified adaptation step and let backporting scale reliably across an advisory's full affected range.

\begin{defn}[Backport Chain]
\label{def:chain}
Let $v_f$ be a fixed version and $v_1 \dots v_n$ be the affected version range for $v_f$.
The backport chain for $v_f$ is the sequence $(v_f, v_n, \dots, v_1)$.
We refer to $v_n$ as the \emph{highest affected version} and to the pair $(v_f, v_n)$ as the \emph{fixed-vulnerable pair}.
\end{defn}

These principles shape the four stages of Figure~\ref{fig:system-overview}, each addressing one challenge from Section~\ref{sec:problem}.
\squaretext{gray}{1}~\autoresearch addresses data quality.
It verifies and enriches the advisory's data, identifies the security-relevant portions of the upstream fix, and resolves the affected ranges into backport chains.
It emits two artifacts that every later stage consumes.
The \emph{vulnerability manifest} holds the verified and enriched advisory data.
The \emph{cross-version map} records how the vulnerable code and its surroundings change across the versions of each chain.
\squaretext{gray}{2}~\autopoc addresses exploit validation.
It generates executable \pocs from the manifest and verifies each against the fixed-vulnerable pair, requiring success on the vulnerable build and failure on the patched build.
The verified \pocs establish which versions are actually exploitable and serve as the \emph{vulnerability oracle} for every candidate backport.
\squaretext{gray}{3}~\autotester addresses regression validation.
It generates regression tests that pin the intended behavior around the vulnerable code path without preserving the vulnerability itself, and verifies that they pass on both builds.
These tests form the \emph{functionality oracle}.
\squaretext{gray}{4}~\autopatch addresses code divergence.
\autopatch first isolates the \emph{minimal security patch} (MSP).
It then adapts the MSP down each chain and accepts a backport only when it clears both oracles.

\begin{defn}[Minimal Security Patch]
\label{def:msp}
Let the upstream fix be the diff between the fixed-vulnerable pair $(v_f, v_n)$ (Def.~\ref{def:chain}), partitioned into hunks $H = \{h_1, \dots, h_m\}$.
Each hunk is either a \emph{security hunk}, one required to block the vulnerability, or a \emph{non-security hunk}.
The minimal security patch $\mathcal{M} \subseteq H$ is the set of security hunks.
\end{defn}

Section~\ref{sec:methodology} details how each stage within \toolname assembles this evidence.

%% file: sections/methodology.tex
\begin{figure*}
    \centering
    \includegraphics[width=1\linewidth]{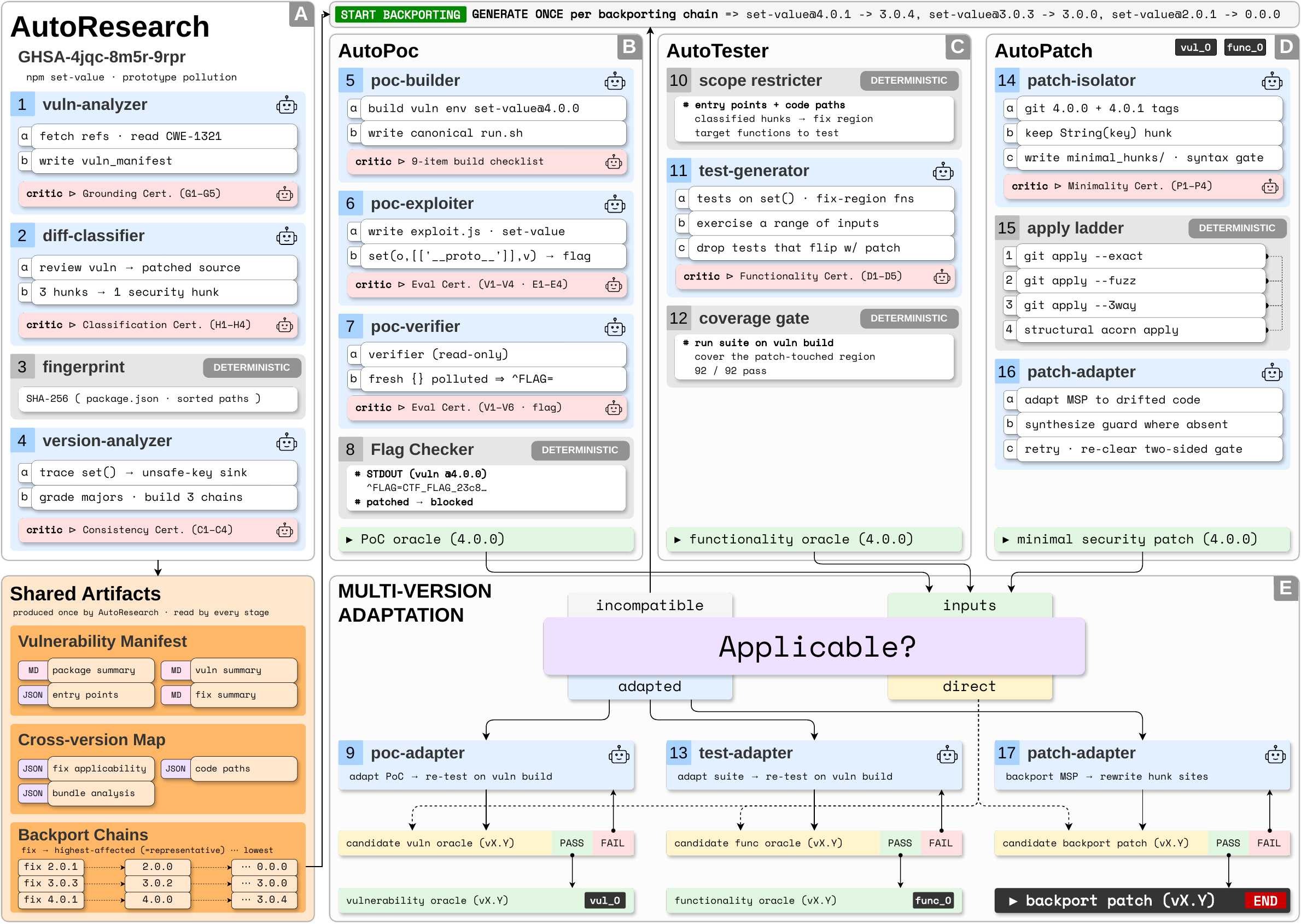}
    \caption{\toolname system design. \texttt{GHSA-4jqc-8m5r-9rpr} is provided as a running example.}
    \label{fig:system-design}
\end{figure*}

\toolname takes a vulnerability advisory and produces a \emph{verified} backport for every affected version, decomposing the task into the four stages of Figure~\ref{fig:system-design}.
\autoresearch grounds the advisory, \autopoc and \autotester build a vulnerability and a functionality oracle, and \autopatch isolates and adapts the fix.
We call a backport verified when an executable exploit can no longer trigger the vulnerability on it and an executable regression suite still passes; this is evidence from two oracles, not a proof, since each oracle witnesses specific behavior rather than the absence of every vulnerable path.
Each stage targets one of the backporting challenges of Section~\ref{sec:problem} while addressing the LLM limitations discussed in Section~\ref{sec:OurApproach}, and deterministic scaffolding (git-apply ladder, fingerprinting, scope bounding, and flag checking) bounds where stochastic agents may act.
We instantiate \toolname for the npm ecosystem and use the advisory \texttt{GHSA-4jqc-8m5r-9rpr} for \texttt{set-value} as a running example throughout this section.

\subsection{\autoresearch}
\label{sec:autoresearch}

Advisory data includes raw information about the package, the vulnerability, and the fix that later stages depend on.
This information is often unreliable, with incorrect version ranges, imprecise summaries, and references that point to external sources whose own PoCs and fix descriptions may be inaccurate.
The goal of \autoresearch is to produce a \emph{vulnerability manifest} that verifies and extends this information and places the advisory in a normalized form the rest of the pipeline consumes.

\subsubsection{Advisory Refinement}
\label{sec:advisory-refinement}
\autoresearch compares the vulnerability the advisory describes against the vulnerability as it appears in source.
Because the vulnerability can manifest differently across versions, and showing an agent every version would overload its context (Section~\ref{sec:OurApproach}), \autoresearch provides the \emph{vuln-analyzer} agent~\squaretext{gray}{1} with the fixed-vulnerable pair (Def.~\ref{def:chain}) for a given backport chain.

The agent also reads the advisory description, CWE, severity, and references.
The references often carry the richest evidence, since they can link issues containing PoCs, vulnerability-fixing commits (VFCs), and pull requests in which the maintainers explain the fix.
A listed reference can link onward to richer evidence than it contains, and valuable sources such as a security write-up's proof of concept may go unlisted in the advisory.
A deep-research step within the vuln-analyzer workflow follows links deeper within a reference and out to related sources to gather this evidence, bounded to a fixed link depth and a curated allowlist of authoritative domains.
For example, the advisory for CVE-2025-55182, a deserialization flaw in React Server Components, listed a third-party repository as its proof-of-concept reference; by reading that repository's issues alongside the upstream fix, deep-research found that the listed exploit did not trigger the vulnerability.
Left unchecked, that reference would have reached \autopoc as ground truth and produced an exploit for the wrong code path.
The end result is the \emph{vulnerability manifest}, which records the package, vulnerability, and fix summaries together with the package's entry points and a structured report of every reference reviewed and its relevance.

\subsubsection{Version Analysis}
\label{sec:version-analysis}

Adapting the patch, PoC, and regression suite to every affected version requires knowing what each version contains and how far it has drifted from the fix.
Analyzing every version individually would be wasteful, since affected versions can be structurally identical and differ in only a few lines.
\toolname therefore fingerprints each version's shape~\squaretext{gray}{3} as a SHA-256 over its \texttt{package.json} entry fields and sorted file paths, and groups versions whose fingerprints match.
The fingerprint is deterministic and content-blind: a shared fingerprint means two versions have the same file layout and package entry structure, not necessarily identical contents.
Reusing a representative's \emph{structure} report across its group is therefore sound, whereas reusing its content-dependent \emph{code-path} and \emph{fix-applicability} reports is a heuristic that assumes versions of identical shape expose the vulnerability alike.
\toolname accepts this gap because the fingerprint captures exactly the structural divergence (Section~\ref{sec:challenge-code-divergence}) that defeats mechanical reuse, and because every reused artifact is later re-checked by execution on its own version, so an unsound reuse surfaces as a failed oracle rather than a silent error.
A cosmetic layout change instead over-fragments a group, wasting analysis effort but never merging versions that truly diverge in shape.
\toolname analyzes one representative per group rather than once per version.

A \emph{diff-classifier} agent~\squaretext{gray}{2} makes a pass over the hunks $H$ of the upstream fix (Def.~\ref{def:msp}) and gives each a preliminary security or non-security label.
When the vulnerability manifest recovered a fixing commit, this classification is scoped to the files that commit touched; when no fixing commit was recovered, the common case for advisories (Section~\ref{sec:challenge-data-quality}), the classifier instead ranges over the full fix diff.
This triage is fast rather than exhaustive, producing only a rough estimate of \msp; the authoritative isolation happens later in \autopatch (Section~\ref{sec:autopatch}).

\myparagraph{Analyzing Each Group}
A \emph{version-analyzer} agent~\squaretext{gray}{4} then examines one group representative, comparing it against the fixed version using the security hunks and the manifest's entry points, and emits three reports.
A \emph{structure} report records how the package is laid out and built, such as its entry fields and whether it ships built artifacts a patch must also reach.
A \emph{code-path} report traces each entry point to the vulnerable sink and confirms the vulnerability is reachable in that group.
A \emph{fix-applicability} report grades how the fix lands as \emph{direct}, \emph{adapted}, or \emph{incompatible}, together with the concrete differences behind the grade.
For \texttt{GHSA-4jqc-8m5r-9rpr}, the analyzer returns \emph{adapted} for \texttt{0.x} and \emph{direct} for the other three groups.
The \texttt{3.x} group is graded \emph{direct} because its representative already defines the guard the fix hardens, \texttt{isValidKey}, so the coercion applies in place.
The \texttt{0.x} group is graded \emph{adapted} for a sharper reason, since its representative lacks \texttt{isValidKey} entirely, so the fix cannot be transplanted and must be synthesized against a different key-splitting path.
\toolname stitches these reports into the \emph{cross-version map}, which tells each later stage which files and code paths the fix must touch in a given version and what must change for it to land.

\subsection{\autopoc}
\label{sec:autopoc}

A backported patch that applies and even builds correctly can still fail to truly block the vulnerability.
A Proof-of-Concept exploit (PoC) that reproduces the vulnerable behavior provides executable evidence that a patch correctly mitigates the stated vulnerability.
\toolname originally had a single agent write this exploit from the advisory, which worked for simple cases but proved brittle across projects and could not handle incorrect affected ranges.
Reliable \poc generation instead has to \emph{build} a working environment, craft an \emph{exploit} that genuinely triggers the vulnerability, and \emph{verify} that the exploit fired while rejecting false positives.
CVE-Genie~\cite{CVEGENIE} covers these three steps with a \emph{build}$\rightarrow$\emph{exploit}$\rightarrow$\emph{verify} chain of agents that reproduces a vulnerability for a particular CVE.
\autopoc extends this chain for backporting, producing a \emph{vulnerability oracle} for each backport: a \poc that fires on a vulnerable version's build and is blocked on the patched version.

\subsubsection{PoC Generation}
\autopoc uses the \emph{vulnerability manifest}, \emph{cross-version map}, and a fixed-vulnerable pair per backport chain, rather than the raw advisory.
A \emph{poc-builder} agent~\squaretext{gray}{5} stands up a working environment for the highest affected version.
Then a \emph{poc-exploiter} agent~\squaretext{gray}{6} writes an attack against it, anchored by the vulnerable path and sink extracted in the vulnerability manifest.
CVE-Genie's \emph{poc-verifier} agent~\squaretext{gray}{7} confirms that the exploit fires on the vulnerable build, but not that it is blocked once the patch is applied, which is the gate \autopoc adds next.

\myparagraph{Verification}
Automated exploit generation is prone to false positives, exploits that satisfy a validator without ever triggering the vulnerability~\cite{POCGEN, CVEGENIE}.
Like CVE-Genie~\cite{CVEGENIE}, \autopoc has an LLM author a flag-based verifier whose embedded flag is read by a deterministic, non-LLM check~\squaretext{gray}{8}.
\autopoc adds a second gate against this, requiring the same exploit that fires on the vulnerable build also be blocked once the patch is applied.
A check is only as honest as the build it runs on, so the harness rather than the agent installs the exact version, applies the patch, and rejects any run where the exploit reinstalled the vulnerable code or the patched code never loaded.
Because the patched build passes when the exploit does not fire, differential guards ensure a pre-existing crash, an infrastructure failure, or a verdict with no executed checks is never taken as a pass.
The result is the \emph{vulnerability oracle} for the highest affected version of the chain, which establishes this exploit is blocked on the patched build rather than that no exploit remains.
When \autopoc cannot build a working exploit for a version, that version carries no vulnerability oracle, and \autopatch reports its backport as unverified rather than certifying it.

The exploit for \texttt{GHSA-4jqc-8m5r-9rpr} wraps an unsafe key in an array to slip past a blocklist an earlier fix added.
On the vulnerable build it pollutes a fresh object and \autopoc captures its flag, while the patched build coerces the key to a string the blocklist rejects, so the object stays clean.
The flag is gated on the observed pollution, so a build that never pollutes cannot pass verification.

\subsubsection{Multi-Version Adaptation}
\label{sec:autopoc-Adaptation}
\autopoc carries each vulnerability oracle across the rest of the affected range for its backport chain rather than re-running the \emph{build}$\rightarrow$\emph{exploit}$\rightarrow$\emph{verify} chain for each version.
It reuses \autoresearch's fingerprint groups (Section~\ref{sec:version-analysis}), validating the oracle once per group representative and adapting it to that group's structural siblings.
As the versions share one vulnerability and differ only in structure, the work left for each is to fit the exploit to its own code, and \autoresearch's \emph{cross-version map} has already graded how far each has drifted.
Where the surface is unchanged (\texttt{direct}), the validated exploit is reused verbatim and a single agent-free harness run certifies the version if it still passes, falling back to adaptation if it does not.
Where references have drifted (\texttt{adapted}), a \emph{poc-adapter} agent~\squaretext{gray}{9} rewrites only what moved, such as import paths, function signatures, and return shapes, and leaves the verified exploit logic intact.
Only when the package has been reorganized enough that the fix no longer maps (\texttt{incompatible}) does \autopoc regenerate a fresh oracle for the group representative and adapt inward to its siblings.
For \texttt{GHSA-4jqc-8m5r-9rpr}, the exploit drives only the package's public entry point, which is stable across the range, so \autopoc reuses it on every version and re-checks the two-sided result rather than regenerating an oracle.

\myparagraph{Advisory Range Correction}
Published advisories routinely misstate the affected version range in both directions, listing \emph{false-positive} releases that are not actually affected and omitting \emph{false-negative} releases that are.
Pinning down either error requires running a real exploit against the release and observing whether it fires.

False positives surface naturally while \autopoc builds its per-version oracles across the affected range.
When an oracle verified on a neighboring affected version fails to trigger on a listed version, and \autoresearch also grades that version structurally incompatible with the vulnerable path, \toolname flags the release as a candidate false positive.
For example, the advisory \texttt{GHSA-vg7j-7cwx-8wgw} for the widely used \texttt{mongoose} package marks much of its release history as affected, yet \toolname's oracle does not fire on 422 of these releases, the kind of over-approximation that motivates \autoresearch (Section~\ref{sec:challenge-data-quality}).

False negatives are harder, since confirming them requires oracles for versions outside the affected range, and testing every such version naively would be prohibitively expensive.
\autopoc instead runs a \emph{differential boundary search} that probes the immediate neighbors of each edge of the affected range.
At the lower edge it walks outward one version at a time while the oracle keeps firing, stopping at the first release it cannot exploit, and that last exploitable release becomes the corrected boundary.
At the upper edge it instead probes only the single release just above the fix: a vulnerability surviving the stated fix usually signals a broken patch rather than a wider range, so \toolname deliberately favors precision over recall here and does not walk further, accepting the risk of missing a true upper-edge false negative.
A clean result confirms the boundary, while an exploit there is flagged as a likely incomplete upstream fix for review rather than widening the range automatically.
For example, the advisory for the \texttt{vm2} sandbox escape (\texttt{GHSA-ffh4-j6h5-pg66}) initially named a single affected release, but \toolname's oracle fired on earlier releases as well, revealing that the escape was present well before the named version; the upstream advisory now reflects this wider range.

\subsection{\autotester}
\label{sec:autotester}

A backported patch that blocks the vulnerability can still break intended functionality within the package.
A regression suite guards against this by checking that a patch preserves existing safe behavior.
However, crafting these tests is subtle because a security regression test must distinguish between the behavior the patch removes and the intended behavior it preserves.
\autotester produces a \emph{functionality oracle} for each backport, a regression suite that passes on both the vulnerable and patched builds.

\subsubsection{Test Generation}
\label{sec:autotester-gen}

A \emph{test-generator} agent~\squaretext{gray}{11} grounded in \autoresearch's artifacts writes tests that exercise the package's public API, and a separate critic agent reviews them.
These tests run against both the package source and the build artifacts it ships, such as npm's \texttt{dist} and \texttt{esm} bundles.
The agent refines the suite over several rounds, guided by the critic's feedback and a coverage gate~\squaretext{gray}{12} that each run reports.

\myparagraph{Focused Generation}
A patch is most likely to perturb behavior near the code it changes (Section~\ref{sec:challenge-regression-val}), so for large packages a \emph{scope restricter}~\squaretext{gray}{10} confines generation to that region rather than the full public API, where exercising every entry point would be prohibitively expensive.
This focusing is gated on package size, taking effect once a package exceeds a configurable file-count threshold, 50 source files in our deployment.
We set the cutoff well above the size of a typical npm package, which carries on the order of 20 source files, so the common case is still tested comprehensively and only the large frameworks above it, where comprehensive generation runs to millions of tokens, are scoped down.
Package sizes are heavily skewed into a cheap majority and a costly tail, so the precise cutoff matters little provided it separates the two regimes, which is why we leave it configurable rather than tuned.
For a focused run, it reads \autoresearch's entry points and traced code paths to find which functions to target and the classified hunks to see what the fix changes, then exercises those functions with a range of inputs to cover the behavior the fix can affect.
Because these are exactly the paths a patch can perturb, a backport that over-restricts the fixed code is caught here; behavior far from the fix is out of scope by design, on the assumption that a minimal security patch leaves it untouched.

\myparagraph{A Patch-Invariant Oracle}
A regression test must stay independent of the fix, so \toolname keeps a test only when it passes on both the vulnerable and the patched build, making the suite \emph{patch-invariant}.
A test that fails on the vulnerable build reflects a pre-existing failure rather than a regression.
A test that asserts on the behavior the patch changes is the opposite hazard, flipping from pass to fail once the fix lands and mistaking the fix for a regression, so the generator and its critic drop it, leaving the suite invariant by construction.
\toolname then verifies this rather than assuming it.
After \autopatch applies the backport, it re-runs the suite on the patched build.
A failure is now unambiguous.
One that appears only after the patch is a genuine regression, while one present on both builds is a pre-existing failure, set aside rather than blamed on the patch.

The functionality oracle for \texttt{GHSA-4jqc-8m5r-9rpr} exercises the package's ordinary key-setting behavior, such as nested writes like \texttt{set(obj, 'a.b.c', v)}.
Its security-adjacent tests use string keys like \texttt{\_\_proto\_\_}, which the blocklist already rejects on both the vulnerable and the patched build.
Each such test therefore yields the same verdict on either side of the fix.
A test built on the array-wrapped key the exploit relies on would instead flip between builds, since that key pollutes before the fix and is rejected after it.
\autotester's critic re-derives the fix and confirms that no test depends on this path, leaving the suite patch-invariant by construction.

\subsubsection{Multi-Version Adaptation}
\label{sec:autotester-adaptation}
\autotester carries the \emph{functionality oracle} across the affected range with the same hierarchical strategy as \autopoc (Section~\ref{sec:autopoc-Adaptation}), driven by the \emph{cross-version map}'s \texttt{direct}, \texttt{adapted}, and \texttt{incompatible} grades.
Where the surface is unchanged (\texttt{direct}), the suite is reused as is and a single harness run certifies the version if it still passes.
Where references have drifted (\texttt{adapted}), a \emph{test-adapter}~\squaretext{gray}{13} subagent rewrites the moved function signatures, import paths, and return shapes that the \emph{cross-version map} records, and leaves the assertions intact.
Holding the assertions fixed is what carries patch-invariance to the adapted version, since the adapter changes only how the suite reaches the API and not the behavior it checks.
Only when the package has been reorganized enough that the suite no longer maps (\texttt{incompatible}) does \autotester regenerate a fresh suite for the group representative and adapt inward to its siblings.
Each adapted suite must pass on its own vulnerable build to become that version's \emph{functionality oracle}, and its patched-build check completes once \autopatch applies that version's patch, as with the representative version.

For \texttt{GHSA-4jqc-8m5r-9rpr}, \texttt{set-value} exposes one stable \texttt{set(obj, path, value)} entry point across all four groups.
The functionality oracle is therefore reused without adaptation: each version is certified by a single harness run, and only its patched-build re-check waits until that version's backport lands.
Had the public surface drifted in some group, the \emph{test-adapter} would rewrite only how the tests call the API, leaving the \texttt{set(obj, 'a.b.c', v)} assertions untouched, so the suite checks the same behavior and returns the same verdict on both builds.

\subsection{\autopatch}
\label{sec:autopatch}

\autopatch produces a verified backport for each affected version of the package.
A natural design has a single agent adapt the fix directly from the vulnerable and patched versions, but we found that such an agent overlooks changes that appear unrelated to the vulnerability and loses track of the original fix on a long chain, leaving the backport incomplete.
\autopatch instead separates patch isolation from patch adaptation and uses three steps for each chain.
(1) It isolates the minimal security patch \msp (Def.~\ref{def:msp}) from the upstream fix.
(2) It adapts \msp into a candidate backport for the highest affected version $v_n$ (Def.~\ref{def:chain}) and accepts it only when the vulnerability oracle blocks the exploit and the functionality oracle still passes.
(3) It carries the verified backport across the remaining versions in the chain.

\subsubsection{Patch Isolation}
\label{sec:autopatch-isolate}
\autopatch reuses the vulnerability manifest and cross-version map that \autoresearch produced, which describe the vulnerability, the package, and the upstream fix.
It works in a dedicated Git workspace that holds two immutable checkouts, the fixed-vulnerable pair $(v_f, v_n)$ (Def.~\ref{def:chain}), which the agents read but never modify.
\autopatch deterministically diffs the pair into the hunks $H$ of the upstream fix (Def.~\ref{def:msp}).

The \textit{patch isolator} agent~\squaretext{gray}{14} labels each hunk in $H$ as a security hunk or a non-security hunk.
The accepted security hunks form \msp, the MSP for the chain, which \autopatch holds as an immutable reference to the original fix.
The non-security hunks cover the refactoring, formatting, and unrelated edits the version bump happens to include.
The agent grounds this classification in \autoresearch's compiled evidence (package, vulnerability, and fix summaries, code paths) and confirms it by exploring the two versions, after which a critic re-examines it against the same evidence and either accepts it or returns it for another round.
Because the classification is not executable, \msp is a textual estimate rather than a proven minimum, and execution during backporting can still reveal it to be incomplete.

Of \texttt{GHSA-4jqc-8m5r-9rpr}'s three chains, we follow the chain for 3.0.3 (affected versions 3.0.0--3.0.2).
The upstream diff bundles README and \texttt{package.json} churn with the fix.
The isolator keeps a single four-line hunk, the \texttt{String(key)} coercion, as \msp.

\subsubsection{Backporting the Highest Affected Version}
\label{sec:autopatch-rep}
The \textit{patch adapter} agent~\squaretext{gray}{16} reshapes \msp to fit the highest affected version $v_n$ (Def.~\ref{def:chain}), whose code might have drifted from the upstream fix.
It works in an editable checkout of $v_n$ and keeps the immutable fixed-vulnerable checkouts available for comparison.
For each security hunk in \msp, the agent produces an \emph{adapted hunk} that fits $v_n$'s code.
The agent never edits \msp itself, so the original fix stays intact for the rest of the chain.

Unlike the hunk classification in isolation, a candidate backport can be executed, so \autopatch verifies it with the two oracles instead of a critic.
The vulnerability oracle must block the exploit on the patched build and the functionality oracle must pass, and the agent revises the candidate and retries whenever either fails.
A failure can reveal that \msp does not fit this version, either because one of its security hunks does not apply or because a hunk that isolation labeled non-security is in fact needed here.
The agent responds by adding or removing hunks from the candidate, never by editing \msp itself, so execution refines the security-relevant set that isolation could only estimate textually.
The accepted adapted hunks form the backport for $v_n$.

One release below the fix, the highest affected version 3.0.2 already defines \texttt{isValidKey}.
The MSP applies as written and clears both oracles.

\subsubsection{Multi-Version Backporting}
\label{sec:autopatch-multiversion}
\autopatch carries the verified backport from $v_n$ toward $v_1$ along the chain (Def.~\ref{def:chain}), using the same hierarchical strategy as \autopoc and \autotester, driven by the cross-version map's \texttt{direct}, \texttt{adapted}, and \texttt{incompatible} grades.

To land the backport on each version, \autopatch climbs an escalating git \texttt{apply ladder}~\squaretext{gray}{15} and takes the first rung that produces a syntactically valid, oracle-verified patch.
The lower, agent-free rungs run \texttt{git apply} with progressively looser matching (exact, fuzzed context, then three-way merge), followed by a structural pass that relocates each hunk by its parsed context and rewrites renamed identifiers; only when all of these fail does the top rung invoke the patch adapter agent~\squaretext{gray}{17}.
Versions whose surface is unchanged (\texttt{direct}) typically clear a mechanical rung and apply without an agent, whereas versions whose references have drifted (\texttt{adapted}) fall through to the patch adapter, which rewrites how each hunk lands in the version, and leaves \msp intact.
Grounding each adaptation in \msp's immutable hunks rather than the backport it starts from is what keeps the fix faithful across a long chain, since an agent that adapts from an already-backported version would otherwise inherit its drift and lose track of the original fix.
Only when the package has been reorganized enough that the backport no longer maps (\texttt{incompatible}) does \autopatch adapt \msp afresh for the group representative and adapt inward to its siblings.

A version's backport is accepted only once that version's vulnerability oracle blocks the exploit and its functionality oracle passes on the patched build, as at $v_n$; a version that never clears both oracles is reported as an unverified backport rather than shipped as a fix, so a failure is surfaced rather than hidden.
Once every version holds a verified backport the chain is complete, and repeating this across all chains yields a verified fix for every affected version.

On 3.0.1 the hunk applies unchanged.
Version 3.0.0 inherits the group's \texttt{direct} grade through its content-blind fingerprint, but \texttt{isValidKey} does not exist there.
The mechanical rungs fail, so the adapter synthesizes the guard and filters the parsed keys through it.
Both oracles accept the result even though it no longer resembles the upstream fix.
The grade routes the attempt; the oracles decide.

\subsubsection{Patch Refinement}
\label{sec:autopatch-refinement}
A backport is accepted once it clears both oracles, but that verdict can rest on incomplete evidence, for instance when a stronger \poc later exposes a residual bypass or a functionality failure surfaces only after the patch lands.
When such a failure appears, \autopatch enters a refinement loop rather than discarding the backport.
It forks a new revision from the failing one, never mutating the accepted record, and runs a bounded fix-and-verify loop guided by the failure.
A surviving exploit directs it to harden the gate, whereas a regression with the exploit still blocked directs it to repair functionality without weakening the security fix.
The loop is fail-closed, so a revision is promoted only when it re-clears both oracles, and an iteration that reproduces the prior patch unchanged is rejected rather than accepted as progress.

%% file: sections/eval.tex
This section evaluates \toolname by answering the following research questions.
\begin{enumerate}[label=\textbf{RQ\arabic*},labelwidth=2em,leftmargin=!,itemsep=0pt,parsep=0pt]
    \item \textit{How effective is \toolname relative to existing backporting techniques?}
    \item \textit{How effective is \toolname at backporting across the npm ecosystem?}
    \item \textit{How much does each component of \toolname contribute to backporting effectiveness?}
\end{enumerate}

\subsection{Experimental Setup}
\label{sec:eval-setup}

We now define the benchmarks, task inputs, comparison tools, and metrics used throughout the evaluation.
For reasons discussed in the corresponding subsections, RQ1 uses BackportBench, while RQ2 and RQ3 use \benchname, a benchmark we created for this paper.
Additionally, RQ1 compares \toolname to two patching tools and two generic agents.
RQ2 and RQ3 only compare to Claude Code, the best-performing baseline from RQ1.

\subsubsection{Benchmarks}
Our evaluation uses two benchmarks:

\myparagraphn{BackportBench}~\cite{BackPortBench} is a repository-level, execution-validated benchmark for backporting.
It is constructed from existing vulnerability-fixing commits and their maintainer backports.
Each task evaluates whether a tool can adapt an existing upstream fix to a historical target version for which a maintainer-written backport is already available.
The released benchmark contains 202 tasks from 12 repositories across three ecosystems: PyPI, Maven, and npm.
Each task consists of an upstream (golden) patch, an unpatched target version, a containerized execution environment, and two test lists: Fail-to-Pass (F2P) tests that encode the expected fix behavior and Pass-to-Pass (P2P) tests that guard against regressions.
A task passes when all F2P and P2P tests pass.

We excluded two types of tasks from BackportBench:
(1) 67 Maven tasks, as \toolname does not currently support Java,
(2) 7 tasks (2 PyPI and 5 npm) for which we could not reproduce a passing result from the provided inputs.
Appendix~\ref{appendix:excluded-tasks} lists each excluded task and its reason.
The resulting BackportBench evaluation set contains 110 PyPI tasks and 18 npm tasks.

\myparagraphn{\benchname}
is a benchmark we created to evaluate npm backporting more broadly than BackportBench supports.
BackportBench constructs tasks from maintainer-written backports, and npm maintainers rarely backport~\cite{PLUMBER}, so its released set contains only 23 npm tasks from four repositories covering 9 CVEs and 11 CWEs.
\benchname instead was created by combining output from a preliminary version of \toolname and significant human manual curation.
For each included task, a human security expert (1)~reviewed the PoC and regression test suite, (2)~edited or replaced incorrect artifacts, (3)~confirmed the final PoC triggers the vulnerable release,  and (4)~confirmed the PoC is blocked by the patched reference release.
The expert also ensured the regression suite exercises package behavior that should be preserved.
This lets \benchname cover \benchmarkAmount npm tasks from 44 repositories, spanning 60 CVEs and 33 CWEs.

Each \benchname task contains: (1)~a vulnerable package release, (2)~a GitHub security advisory (GHSA), (3)~a containerized execution environment, (4)~a proof-of-concept exploit, (5)~a regression test suite, and (6)~a reference backport that blocks the PoC and passes the regression suite.
\benchname resolves each GHSA affected range to stable npm releases, groups them by affected major version, and selects the lowest, median, and highest affected release in each major line when those releases exist.
This design gives \benchname greater affected-range coverage than BackportBench while keeping the benchmark small enough for human review.
\benchname evaluates 6.5 versions per advisory on average versus 1.8 for BackportBench.
It covers 19.6\% of affected ranges on average versus 7.5\%.

\subsubsection{Comparison Tools}

We compare \toolname against four baselines.
All tools run in isolated containers with identical CPU, memory, and wall-clock limits.
RQ1 experiments use Claude Opus 4.7\cite{OPUS4-7} and RQ2 and RQ3 experiments use Claude Opus 4.6~\cite{OPUS4-6}, the strongest coding models available at the time of each experiment.

\myparagraphn{Claude Code}~\cite{CLAUDECODE} is Anthropic's production agentic coding tool. 
Since \toolname is implemented on top of Claude Code, this baseline isolates the contribution of \toolname's task decomposition, artifact contracts, critics, and verification workflow beyond the underlying coding agent.

\myparagraphn{mini-swe-agent}~\cite{SWEAGENT} is the maintained successor to SWE-agent, an agent-computer-interface system designed for repository-level automated repair, where an LLM navigates a codebase, edits files, and runs tests to produce patches for GitHub/SWE-bench-style issues.
mini-swe-agent preserves this repair-oriented workflow while reducing the interface to a lightweight bash-only tool.
We use the official \texttt{swebench} configuration distributed with mini-swe-agent.

\myparagraphn{MagentLess}~\cite{MAGENTLESS} is a multilingual adaptation of Agentless~\cite{AGENTLESS}, an llm-based procedural repair tool that localizes relevant files and functions from a repository and generates edit operations.
We adapt MagentLess for backporting by providing the upstream patch as the problem description and converting its edit output into a unified diff for evaluation.

\myparagraphn{PortGPT}~\cite{PORTGPT} is the state-of-the-art academic agentic backporting tool that uses a two-stage hunk-adaptation and patch-validation workflow with tools for code viewing, Git-history tracing, hunk application, and compile feedback.
We extended PortGPT's CompileTest stage with three language-native parse-level validators: \texttt{node --check} (JavaScript), \texttt{tsc --noEmit} (TypeScript), and \texttt{py\_compile} (Python).

\myparagraph{Excluded Systems}
We considered direct comparisons with prior automated backporting systems, including FixMorph~\cite{FixMorph}, TSBPORT~\cite{TSBPORT}, SKYPORT~\cite{SKYPORT}, and Mystique~\cite{Mystique}.
These systems are not straightforward to adapt to Python and JavaScript, because their core algorithms are coupled to language-specific analyses.
Porting them would therefore require reimplementing front-end parsers, symbol analysis, patch-type rules, and validation harnesses for Python and JavaScript, which would constitute a new system rather than a faithful baseline adaptation.

\subsection{RQ1: Comparison to Existing Techniques}
\label{sec:RQ1}

We compared \toolname against the baseline tools using BackportBench, as it provides outside validation of our tool.
We ran each tool ten times per task and measured both single-run success and reliability across runs rather than judging it on a single outcome.
Multiple runs are necessary to measure reliability rather than best-case success due to the stochastic nature of LLM agents~\cite{CVEGENIE, TAUBENCH}.

\myparagraph{Repeated-run metrics}
Let \(c_i\) be the number of successful runs for task \(i\) out of \(n=10\) runs.
Following \(\tau\)-bench~\cite{TAUBENCH}, we report \(\text{pass}^{1}=\mathbb{E}_i[c_i/n]\) as expected single-run success.
We report \(\text{pass}^{k}=\mathbb{E}_i\left[\binom{c_i}{k}/\binom{n}{k}\right]\), which estimates the probability that all \(k\) sampled runs solve the task.
We report \(\text{pass@}k=1-\mathbb{E}_i\left[\binom{n-c_i}{k}/\binom{n}{k}\right]\), which estimates the probability that at least one of \(k\) sampled runs solves the task.
We use \(\text{pass}^{10}\) to represent reliability and \(\text{pass@}10\) as best-of-ten recall.

\myparagraph{Benchmark Execution Settings}
We evaluated BackportBench in two settings.
(1)~In the \defaultBackportbench setting, each tool receives the benchmark-provided upstream MSP to adapt to its target version. 
This setting implicitly isolates the patch requiring adaptation, because the tool does not need to infer the MSP from the patched release.
We also report a \texttt{git apply} baseline, which applies the provided MSP directly without any adaptation.
(2)~In the \repoBackportbench setting, each tool receives directories with the vulnerable and patched versions along with the advisory metadata.
For this setting, we remove any fixing-commit URLs to ensure the evaluated tool recovers the security fix from the difference between the two versions.
This setting requires each tool to isolate the MSP itself before adapting it.
We only evaluate \toolname, Claude Code, and mini-swe-agent in this setting, as
PortGPT and MagentLess require the MSP as input and cannot infer it. 

\subsubsection{Patch Adaptation (with MSP)}

\input{evaluationTables/backportbench_pooled}
\toolname exceeds every baseline on every metric (Table~\ref{tab:backportbench-pooled}).
The margins are small because this setting is nearly saturated.
\texttt{git apply} alone resolves 40.6\% of tasks, and both general agents exceed 97\% \passhat{1}.
The repeated-run metrics expose differences that \passhat{1} hides.
mini-swe-agent reaches 97.3\% \passhat{1} but drops to 89.8\% \passhat{10}, so its single-run success does not survive repeated trials.
\toolname holds the highest \passhat{10} at 98.4\%, 3.1 points above Claude Code, and is the only tool to solve every task within ten runs (100\% \passat{10}).

\takeaway{With the MSP provided, single-run success of BackportBench is saturated and reliability is what separates tools.
}
\subsubsection{Patch Isolation and Adaptation (without MSP)}

The \defaultBackportbench setting hands each tool the MSP, which advisories rarely provide. 
This setting withholds it.
A tool must first recover the fix from the full difference between the vulnerable and patched versions before adapting it.

\toolname is nearly unchanged between the two settings, falling 0.8 points under \passhat{1} pooled across both ecosystems, while Claude Code falls 13.2 points and mini-swe-agent falls 17.3 (Table~\ref{tab:backportbench-pooled}).
The drops are consistent across npm and PyPI (Table~\ref{tab:backportbench-repo}).
\toolname's worst case across ten runs exceeds every baseline's best case on both ecosystems, with a \passhat{10} of 94.4\% on npm and 95.5\% on PyPI against a best baseline \passat{10} of 88.9\% and 91.8\%.

\myparagraph{Observations}
The drop in performance does not come from running out of resources.
Every mini-swe-agent and Claude Code failure we examined stopped on its own, several costing less than the corresponding run with the MSP provided.
Neither agent builds a check of its own, so nothing in the loop contradicts a confident but incomplete fix.

Two failure shapes account for the lost tasks, both from reading the fix off the patched version.
In the first, the agent ports too little.
On \texttt{django\_261}, both agents remove the AJAX bypass named in the advisory but miss the compensating change that lets legitimate callers satisfy the check through a request header, which is the behavior the test verifies.
In the second, the agent ports too much.
On \texttt{django\_193}, both agents transplant host-allowlist machinery from a later major version onto the older target and break existing behavior.

On six tasks, both agents fail every run while \toolname succeeds on all ten, and each fix requires a hunk in a file the advisory never names.
The agents scope their search to the symbols the advisory mentions, so an incomplete advisory leaves them blind to a needed hunk.
\toolname classifies every hunk in the full version diff and gates each patch on both oracles, recovering the hunks an advisory omits and rejecting the over-wide ports that break behavior.

\takeaway{Withholding the MSP costs Claude Code three times more reliability than recall, 22.7 points of \passhat{10} against 7.8 of \passat{10}.
\toolname trades minutes of extra runtime for substantially fewer unreliable or incomplete backports.}

\subsection{RQ2: Backporting Across npm}
\label{sec:eval:npm}

\input{evaluationTables/CVEPatchBench-Comparison}

As noted in Section~\ref{sec:eval-setup}, BackportBench has relatively few npm backporting tasks, while \benchname has \benchmarkAmount.
The fact that npm maintainers rarely backport~\cite{PLUMBER} makes it a valuable target for novel backporting tools.
We now compare \toolname to Claude Code (the best approach in RQ1) using \benchname.

\myparagraph{Results}
As shown in Table~\ref{tab:cvepatchbench-ablation}, \toolname{} scores higher than Claude Code on every metric: by 16.2 percentage points for \passhat{1}, 20.2 for \passhat{3}, and 13.0 for \passat{3}.
Across the three runs, \toolname{}'s worst case (\passhat{3}) is 10.7 percentage points better than Claude Code's best case (\passat{3}).
\toolname{} also has much smaller variance across the three runs: 2.3 percentage points vs Claude Code's 9.5.

\myparagraph{Observations}
We examined nine tasks that Claude Code failed on.
While the patches appear plausible and apply cleanly, they failed due to reasoning gaps.
Eight of the nine patches pass the \benchname regression suite, but the exploit still works.
Without a vulnerability oracle, Claude Code left an exploitable path open and the PoC still succeeded.
The ninth patch ports the upstream fix that blocks the exploit but changes call behavior in a way that fails \benchname's regression suite.
Without a functionality oracle, Claude Code over-reasoned about the exploit and broke functionality.

\myparagraph{Timing}
\toolname is slower than Claude Code on \benchname, averaging 627 seconds per advisory against 313 (Table~\ref{tab:cvepatchbench-ablation}).
This is the expected cost of \toolname's verification-oriented design.
It not only drafts a patch but also isolates the security-relevant change and accepts a candidate only after the exploit and regression oracles validate it.
Relative to Claude Code, \toolname spends about five additional minutes per advisory and improves \passhat{3} by 20.2 points and \passat{3} by 13.0 points.
The added time therefore buys reliability rather than raw throughput.
For security backporting, where a plausible patch can still leave the vulnerability exploitable or break existing behavior, this tradeoff is deliberate.

\takeaway{On \benchname's 393 npm backporting tasks, the reliability gap between \toolname and generic agents persists.
\toolname is slower, but the additional verification work preserves both reliability and recall.
Compared with Claude Code, it improves \passhat{3} by 20.2 points and \passat{3} by 13.0 points.}

\subsection{RQ3: Ablation Study}
We now evaluate the contribution of the \autotester and \autopoc components to \toolname's design.
These components are designed to provide reliability through the use of functionality and vulnerability oracles, respectively.

\myparagraph{Results}
As shown in Table~\ref{tab:cvepatchbench-ablation}, removing either oracle degrades every metric.
The vulnerability oracle contributes the most.
Removing it costs 8.5 percentage points of \passhat{1} and 9.6 of \passhat{3}, against 4.0 and 3.8 for the functionality oracle.
Removing both costs 9.8 percentage points of \passhat{1}, less than the 12.5 point sum of the individual losses.
This demonstrates that the oracles overlap on some failures while each catches defects the other misses.
The oracle-free variant still exceeds Claude Code by 6.4 percentage points of \passhat{1} and 7.8 of \passhat{3}, isolating the contribution of decomposition and \autoresearch's artifacts.

\myparagraph{Observations}
Claude Code also fails on the tasks the ablation variants leave unsolved.
On these tasks the first candidate patch, from \autopatch and Claude Code alike, is plausible, often a textbook mitigation, yet carries a defect subtle enough to survive review.
Two tasks, one per oracle, illustrate the pattern.
In \texttt{deep-object-diff} (CVE-2022-41713), the first patch creates a diff with \texttt{Object.create(null)}, a standard prototype-pollution defense that blocks five of the six exploit vectors.
On a null-prototype object the \texttt{\_\_proto\_\_} write lands as an own property, the remaining vector reads the planted value back, and the exploit still fires on the patched build.
In \texttt{global-modules-path} (CVE-2022-21191), the first patch adopts the upstream rewrite of \texttt{execSync} into \texttt{spawnSync}, a stronger fix that fully blocks the \poc.
The regression suite drives the package through a mocked \texttt{execSync}, the rewrite bypasses the mock, and the suite fails on the patched build.
Neither error is a localization failure.
Such defects are hard to catch by review alone.

\takeaway{Both verification components improve \toolname's reliability, with \autopoc contributing the largest single gain.
Relative to the oracle-free variant, the full system improves \passhat{1} by 9.8 points and \passhat{3} by 12.4 points, at a cost of 229 additional seconds per advisory.}

\subsection{Limitations}
\label{Limitations}
\myparagraph{Testing-data contamination}
Since BackportBench uses known fixes for its backporting tasks, it is possible that the model used (Opus 4.7) was trained on both the vulnerable packages and the published CVE descriptions in our test set, so the model may recall fix locations rather than derive them, inflating apparent performance on advisories disclosed before its training cutoff.
We provide another benchmark for evaluation, \benchname, which is based on new backports that are not listed in an advisory and were created after the model's training cutoff.

\myparagraph{Ecosystem coverage}
Our quantitative evaluation mostly focuses on npm as \toolname has the strongest support for this ecosystem.
We have support for PyPI and early support for other ecosystems but we treat cross-ecosystem generalization as an empirical claim requiring its own study.

\myparagraph{GHSA as the main advisory source}
\toolname primarily consumes GHSA as its advisory source (Section~\ref{sec:autoresearch}).
\toolname has not been evaluated on its ability to ingest vulnerabilities indexed in OSV, ecosystem feeds, or vendor bulletins.
However, extending ingestion is straightforward in principle, since the pipeline depends on a normalized manifest rather than the GHSA shape.

%% file: evaluationTables/backportbench_pooled.tex
\begin{table}[t]
  \centering
  \caption{BackportBench results pooled across npm and PyPI.
    Time is mean wall-clock seconds per task, weighted by task count (110 PyPI, 18 npm).}
  \label{tab:backportbench-pooled}
  \setlength{\tabcolsep}{6pt}
  \begin{tabular}{lcccc}
    \toprule
    Tool & \(\text{pass}^{1}\) & \(\text{pass}^{10}\) & \(\text{pass@}10\) & Time (s) \\
    \midrule
    \multicolumn{5}{l}{\emph{\defaultBackportbench} (with MSP)} \\
    \addlinespace[2pt]
    \toolname{}        & \textbf{99.6\%} & \textbf{98.4\%} & \textbf{100.0\%} & 271 \\
    Claude Code        & 97.4\%          & 95.3\%          & 99.2\%           & 174 \\
    mini-swe-agent     & 97.3\%          & 89.8\%          & 99.2\%           & 60 \\
    PortGPT            & 89.0\%          & 80.5\%          & 93.0\%           & 71 \\
    MagentLess         & 63.6\%          & 50.8\%          & 75.0\%           & 18 \\
    \texttt{git apply} & 40.6\%          & --              & --               & 0.02 \\
    \midrule
    \multicolumn{5}{l}{\repoBackportbench (without MSP)} \\
    \addlinespace[2pt]
    \toolname{}        & \textbf{98.8\%} & \textbf{95.3\%} & \textbf{100.0\%} & 323 \\
    Claude Code        & 84.2\%          & 72.6\%          & 91.4\%           & 179 \\
    mini-swe-agent     & 80.0\%          & 67.2\%          & 90.6\%           & 77 \\
    \bottomrule
  \end{tabular}
\end{table}

%% file: evaluationTables/CVEPatchBench-Comparison.tex
\begin{table}[t]
\centering
\caption{\benchname{} results across 393 tasks with three runs per task.
Each ablation removes the named oracle from \toolname{}.
Time is mean wall-clock seconds per advisory.}
\label{tab:cvepatchbench-ablation}
\setlength{\tabcolsep}{4.5pt}
\begin{tabular}{lcccc}
\toprule
Condition & \(\text{pass}^{1}\) & \(\text{pass}^{3}\) & \(\text{pass@}3\) & Time (s) \\
\midrule
\toolname{} & \textbf{92.9\%} & \textbf{91.3\%} & \textbf{93.6\%} & 627 \\
\midrule
\quad w/o \autotester{} & 88.9\% & 87.5\% & 90.1\% & 498 \\
\quad w/o \autopoc{}    & 84.4\% & 81.7\% & 86.8\% & 524 \\
\quad w/o both          & 83.1\% & 78.9\% & 86.8\% & 398 \\
\midrule
Claude Code & 76.7\% & 71.1\% & 80.6\% & 313 \\
\bottomrule
\end{tabular}
\end{table}

%% file: sections/deployment.tex
A backport shipped to users must also be reviewable, composable with other fixes for the same package version, and delivered in a form downstream users can adopt without changing their dependency graph.
In production at \company\companyappos, \toolname has generated over 5{,}000 verified backported patches~\cite{PUBLISHEDPATCHES} for 169 high- and critical-severity CVEs.

\myparagraph{Human-in-the-Loop Certification}
\label{sec:deploy-certification}
The vulnerability and functionality oracles provide strong evidence, but the agents producing them are inherently stochastic (Section~\ref{sec:OurApproach}).
Therefore, every patch passes a human certification step before release.
A security engineer reviews the upstream fix analysis, isolated security hunks, per-version PoC verdicts, regression-test results, and the agent rationale behind each artifact.
Review thus checks executable evidence and backtraces decisions rather than reconstructing each backport manually, allowing engineers to efficiently process \toolname's output.
When review finds a defect, the patch is withheld, the failing case re-enters the refinement loop (Section~\ref{sec:autopatch-refinement}) with structured feedback, and the concrete error is distilled into reusable guidance so that later runs avoid it.

\myparagraph{Patch Delivery}
\toolname publishes each verified patch with a signed manifest identifying the package PURL, target version, advisory, severity, changed files, and the associated PoC, test, and review artifacts.
The manifest gives scanners, package-manager plugins, and CI systems an integration boundary while preserving provenance back to the certification evidence.

\myparagraph{Multi-CVE Merging}
A single package version can be affected by multiple CVEs, yet patches are typically only generated for a single vulnerability.
Simply composing multiple patches for a given version can cause issues.
One patch may edit the same code as another or invalidate its assumptions.
\toolname therefore composes verified patches targeting the same package version into a single unified patch, preserves the provenance of each constituent advisory, and re-certifies the merged artifact against the accumulated PoCs and regression suites.
The result is one drop-in patch per version, addressing all relevant CVEs.

\myparagraph{Advisory Correction}
\autopoc verifies an exploit against every version in an advisory's range, discovering per version whether the vulnerability is present (Section~\ref{sec:autopoc-Adaptation}).
A security engineer reviews each flagged version against this evidence and files each confirmed case as a pull request to the GitHub Advisory Database to correct the affected range.
The advisory for the \texttt{vm2} sandbox escape (\texttt{GHSA-ffh4-j6h5-pg66}) had been pinned to the single version its reporter tested, while \toolname's per-version verdicts marked many earlier versions equally exploitable, and the reviewed correction widened the range to cover them.
A single \texttt{mongoose} advisory (\texttt{GHSA-m7xq-9374-9rvx}) was corrected in both directions at once.
Versions below \texttt{3.6.0-rc0}, where the vulnerable \texttt{match} option of \texttt{populate} did not yet exist, were not exploitable and were dropped, and one later version on the 5.x line that had backported the fix was excluded as well.
Excluding that release would have orphaned the still-vulnerable 6.x line, so the same review added a new range to re-cover it.

%% file: sections/conclusion.tex
We presented \toolname, which backports an advisory's fix to every affected version and ships each backport with executable evidence: a PoC the patch blocks and a regression suite it passes.
By decomposing backporting into independently verified stages and letting execution, not the model, accept each patch, \toolname resolves 95.3\% of BackportBench tasks across all ten runs, 22.7 points above the strongest baseline, and 91.3\% of CVEPatchBench's 393 tasks across all three runs.
In production it has generated over 5,000 verified patches for 169 CVEs and corrected 23 upstream advisories.

%% file: sections/acknowledgements.tex
We thank Jordan Harband and Marvin Fleischer for their expert security review of \toolname's published patches and for feedback on its PoCs, regression tests, and backports that helped refine the system.
We also thank Husain Sharaf for help crafting the architectural figures for \toolname.

%% file: sections/appendix.tex
\pagebreak
\section{Additional Figures and Tables}
Table~\ref{tab:backportbench-pooled} reports BackportBench pooled across npm and PyPI.
We pool because npm contributes only 18 of the 128 tasks, where a single task shifts its rate by 5.6 points, leaving the per-ecosystem npm numbers coarse.
Table~\ref{tab:backportbench-default} and Table~\ref{tab:backportbench-repo} isolate the with-MSP and without-MSP settings by ecosystem.

\input{evaluationTables/backportbench_repo}
\input{evaluationTables/backportbench_default}

\section{Excluded BackportBench Tasks}
\label{appendix:excluded-tasks}

This appendix details the 7 BackportBench tasks we exclude from our evaluation (Section~\ref{sec:eval-setup}), we omit the 67 Maven tasks as \toolname does not have full support of that ecosystem yet.
We confirmed each case by execution inside the task's released container.
On six tasks, the benchmark's gold patch passes while a patch built faithfully from the task's provided inputs cannot, so the outcome is independent of the evaluated tool.
We will share reproduction details for each task with the BackportBench authors.

\myparagraph{Provided patch omits required changes}
On \texttt{vite\_31} and \texttt{vite\_37} (CVE-2024-31207), the scored fail-to-pass tests require three fixture files that the provided upstream patch omits.
On \texttt{vite\_50} (CVE-2024-23331), the scored test probes an HTML fixture element that the provided patch omits.
On \texttt{socket.io-parser\_71} and \texttt{socket.io-parser\_72} (CVE-2023-32695), the provided patch reflects the 4.x source commit and omits the error-to-throw change that the maintainer's 3.4.3 and 3.3.4 backports and the scored test include.
A patch limited to the provided changes therefore cannot satisfy the scored tests on these five tasks.

\myparagraph{Formatting-sensitive verdict}
On \texttt{django\_650} (CVE-2022-28347), the scored test asserts the exact error-message text of the maintainer's 2.2.28 backport.
The provided upstream patch uses a different string format, so functionally identical patches fail on formatting alone, while the test that exercises the vulnerability passes.

\myparagraph{Harness instability}
On \texttt{django\_211} (CVE-2022-28346), the baseline test run crashes inside the container, truncating its log and recording seven passing regression tests as failures.

%% file: evaluationTables/backportbench_repo.tex
\begin{table}[h]
\centering
\caption{BackportBench without MSP}
\label{tab:backportbench-repo}
\begin{tabular}{llccc}
\toprule
Ecosystem & Tool & \(\text{pass}^{1}\) & \(\text{pass}^{10}\) & \(\text{pass@}10\) \\
\midrule
\multirow{3}{*}{PyPI}
& \toolname{} & \textbf{98.7\%} & \textbf{95.5\%} & \textbf{100.0\%} \\
& Claude Code & 84.5\% & 71.8\% & 91.8\% \\
& mini-swe-agent & 79.8\% & 69.1\% & 90.9\% \\
\midrule
\multirow{3}{*}{npm}
& \toolname{} & \textbf{98.9\%} & \textbf{94.4\%} & \textbf{100.0\%} \\
& Claude Code & 82.8\% & 77.8\% & 88.9\% \\
& mini-swe-agent & 81.1\% & 55.6\% & 88.9\% \\
\bottomrule
\end{tabular}
\end{table}

%% file: evaluationTables/backportbench_default.tex
\begin{table}[h]
\centering
\caption{BackportBench with MSP provided}
\label{tab:backportbench-default}
\begin{tabular}{llccc}
\toprule
Ecosystem & Tool & \(\text{pass}^{1}\) & \(\text{pass}^{10}\) & \(\text{pass@}10\) \\
\midrule
\multirow{6}{*}{PyPI}
& \toolname{} & \textbf{99.6\%} & \textbf{99.1\%} & \textbf{100.0\%} \\
& Claude Code & 97.5\% & 95.5\% & 99.1\% \\
& mini-swe-agent & 97.3\% & 90.9\% & 99.1\% \\
& PortGPT & 89.9\% & 80.0\% & 94.5\% \\
& MagentLess & 67.5\% & 55.5\% & 79.1\% \\
& \texttt{git apply} & 40.0\% & -- & -- \\
\midrule
\multirow{6}{*}{npm}
& \toolname{} & \textbf{99.4\%} & \textbf{94.4\%} & \textbf{100.0\%} \\
& Claude Code & 97.2\% & \textbf{94.4\%} & \textbf{100.0\%} \\
& mini-swe-agent & 97.8\% & 83.3\% & \textbf{100.0\%} \\
& PortGPT & 83.3\% & 83.3\% & 83.3\% \\
& MagentLess & 39.4\% & 22.2\% & 50.0\% \\
& \texttt{git apply} & 44.4\% & -- & -- \\
\bottomrule
\end{tabular}
\end{table}